\documentclass[aps,showpacs,superscriptaddress,prl,preprint]{revtex4-1}
\usepackage{epsfig}
\usepackage{amsmath}
\usepackage{hyperref}

\begin{document}
\title{Low-energy peak structure in strong-field ionization by mid-infrared laser-pulses: two-dimensional focusing by the atomic potential}
\author{Christoph Lemell}
\affiliation{Institute for Theoretical Physics, Vienna University of Technology, Wiedner Hauptstr.\ 8-10, 
A-1040 Vienna, Austria, EU}
\author{Kostantinos I. Dimitriou}
\affiliation{Department of Physical Science and Applications, Hellenic Army Academy, Vari, Greece, EU}
\affiliation{Dept. of Informatics and Computer Science, Tech. Inst. of Lamia, Greece, EU}
\affiliation{National Hellenic Research Foundation, Inst. of Theoretical and Physical Chemistry, Athens, Greece, EU}
\author{Xiao-Ming Tong}
\affiliation{Institute of Materials Science, University of Tsukuba, Ibaraki 305-8573, Japan}
\author{Stefan Nagele} 
\affiliation{Institute for Theoretical Physics, Vienna University of Technology, Wiedner Hauptstr.\ 8-10, 
A-1040 Vienna, Austria, EU}
\author{Daniil V. Kartashov}
\affiliation{Photonics Insitute, Vienna University of Technology, Gusshausstr.\ 138, A-1040 Vienna, Austria, EU}
\author{Joachim Burgd\"orfer}
\affiliation{Institute for Theoretical Physics, Vienna University of Technology, Wiedner Hauptstr.\ 8-10, 
A-1040 Vienna, Austria, EU}
\author{Stefanie Gr\"afe} 
\affiliation{Institute for Theoretical Physics, Vienna University of Technology, Wiedner Hauptstr.\ 8-10, 
A-1040 Vienna, Austria, EU}

\date{\today}

\begin{abstract}
We analyze the formation of the low-energy structure (LES) in above-threshold ionization spectra first observed by Quan et al.\ \cite{quan09} and Blaga et al.\ \cite{blaga09} using both quasi-classical and quantum approaches. We show this structure to be largely classical in origin resulting from a two-dimensional focusing in the energy-angular momentum plane of the strong-field dynamics in the presence of the atomic potential. The peak at low energy is strongly correlated with high angular momenta of the photoelectron. Quantum simulations confirm this scenario. Resulting parameter dependences agree with experimental findings \cite{quan09,blaga09} and, in part, with other simulations \cite{liu10,yan10,kast11}.
\end{abstract}
\maketitle

One of the important fundamental processes of strong-field laser-matter interaction is above-threshold ionization (ATI), a process generally considered to be well understood \cite{becker02,keldysh64,faisal73}. The workhorse for the description of laser-matter interaction, the strong-field approximation (SFA), is expected to be valid in the regime of small Keldysh parameters $\gamma=(I_p/2U_p)^{1/2}$, where $I_p$ is the ionization potential (binding energy) of the electron to be ionized and $U_p=F_0^2/4\omega^2$ the ponderomotive energy associated with the free quiver motion in the laser field with amplitude $\alpha=F_0/\omega^2$. Small $\gamma\ll 1$ corresponds to large field strength $F_0$ and low frequency $\omega$. In this so-called tunneling regime the SFA (and its semiclassical incarnation, the ``simple-man's model'' (SMM) \cite{lew94}) should work well as the influence of the atomic Coulomb potential is reduced to a weak perturbation. It thus came as a major surprise when for short mid-infrared (mid-IR) laser pulses $\lambda > 1600$ nm and intensities of $I\sim 10^{14}$ W/cm$^2$ ($\gamma \ll 1$) an unexpected peak-like low-energy structure (LES) was found in ATI spectra \cite{quan09,blaga09,catoire09} in contradiction to the SFA featuring a smooth photoelectron spectrum. Therefore, the LES was termed an ``ionization surprise'' \cite{faisal09}.

First experimental studies revealed several characteristic features of the LES providing important clues as to its origin. The LES was found to be universally present irrespective of the ionized atomic or molecular target in the energy range between 1 to 10 eV, its peak position and width primarily depend on the laser parameters and are approximately proportional to the ponderomotive potential $U_P \propto \gamma^{-2}$, and the LES is absent for circularly polarized laser fields. The latter finding seems to indicate that re-scattering, more precisely, revisiting of the ionic core by the electron plays a significant role. Consequently, longitudinal (i.e.\ along the laser-polarization axis) and transverse Coulomb scattering were invoked in theoretical explanations of the LES which was found to be present in the full numerical solution of the time-dependent Schr\"odinger equation (TDSE) \cite{blaga09}. Meanwhile, several models have been proposed invoking different and, in part, contradictory notions. While the coupling between the longitudinal and transverse degrees of freedom by Coulomb scattering was emphasized in \cite{liu10,yan10}, the appearance of the LES was demonstrated in a strictly one-dimensional model in the absence of a transverse degree of freedom \cite{kast11}. Furthermore, an LES feature was also found for screened (short-ranged) model potentials \cite{blaga09} raising questions as to the role of long-range Coulomb scattering and focusing.

In the present paper, we relate the LES to a classical two-dimensional focusing in phase space which gives rise to strong correlation between the energy $E$ and the angular momentum $L$ of the LES electrons since for small $\gamma$ or large $U_p$ (compared to the discrete level spacing) a close classical-quantum correspondence is to be expected. We perform quasi-classical simulations using the classical trajectory Monte-Carlo (CTMC) method including tunneling \cite{dim04} which is validated by full TDSE simulations confirming the $E-L$ correlation. We deduce laser-parameter dependences of the LES consistent with experiment \cite{quan09,blaga09}. Moreover, we identify a pronounced carrier-envelope phase (CEP) dependence of the LES which may open another route towards experimentally monitoring the CEP of mid-infrared pulses. Atomic units are used throughout this paper unless otherwise stated.

Our CTMC simulation employs a standard adaptive step-size Runge-Kutta propagator with initial conditions chosen following \cite{delone91}: the starting coordinate on the polarization axis ($x_0=y_0=0$) is given by the tunnel exit in the combined Coulomb and laser fields,
\begin{equation}
z_0=\frac{I_p/F_0+\sqrt{(I_p/F_0)^2-4/F_0}}{2}\approx \frac{I_p}{F_0}\, .\label{eq1}
\end{equation}
Both longitudinal ($p_\|$) and transverse ($p_\perp$) momentum distributions at the tunnel exit are Gaussian distributed centered at $p_{\|,\perp}=0$ \cite{delone91}. While the initial longitudinal momentum distribution turns out to be unimportant for the LES (one could set $p_\|=0$ without significant change of the results), the width of the transverse momentum distribution
\begin{equation}
\bar{p}_\perp=\sqrt{\frac{F_0}{\sqrt{2I_p}}}\label{eq2}
\end{equation}
is key to sampling the relevant phase-space region within which focusing of the LES occurs. Typically, $10^6$ initial conditions from the distribution are sampled and are propagated in the combined laser and atomic Coulomb fields. At the end of the pulse the asymptotic momentum distribution is determined analytically by propagating along Kepler orbits \cite{arbo06}. For the latter we also allow  exponential screening to probe for finite-range effects.

The classical phase-space analysis is checked against full 3D quantum-dynamics simulations. The TDSE is solved by discretizing the coordinate space in a pseudospectral grid and propagating the wave function by the split-operation method in the energy representation \cite{tong97}. When the time-dependent wave function in space reaches the outer region where the atomic potential becomes negligible compared to the kinetic energy, we project the outer region wave function on Volkov states to obtain the momentum distribution \cite{tong06}. This method allows for long-time propagation of the wavepacket without encountering unphysical reflections at the boundary.

\begin{figure}
\centerline{\epsfig{file=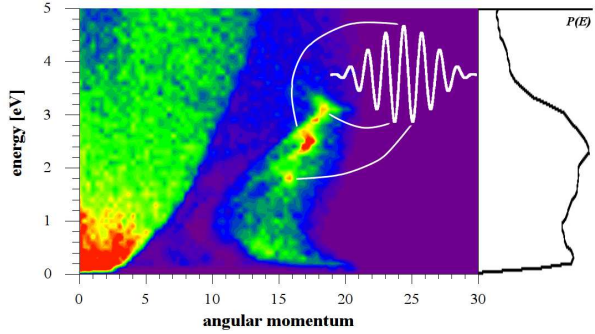,width=8cm}}
\caption{(Color online) Distribution of ionized electrons in the energy ($E$) and angular momentum ($L$) plane observed within an angular cone of $\theta_c=\pm 10^\circ$ around the polarization axis. The island at large values of $L$ is the origin of the LES as demonstrated by the projection onto the energy axis. Inset: electric field of laser pulse (2200 nm, $10^{14}$ W/cm$^2$, 8 cycles)}
\label{fig1}
\end{figure}

\begin{figure}
\centerline{\epsfig{file=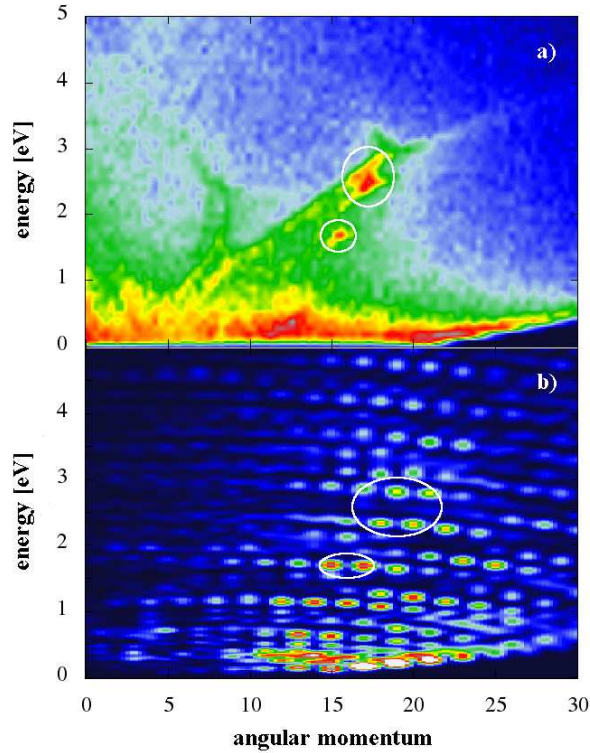,width=8cm}}
\caption{(Color online) Comparison between classical (a) and quantum (b) $E-L$ distributions for angle-integrated emission (same laser parameters as in Fig.\ \ref{fig1}).}
\label{fig2}
\end{figure}

As one typical reference laser pulse we choose a cosine-like pulse shape (carrier-envelope phase $\phi_{CEP}=0$) with $\sin^2$-envelope for the electric field, a wavelength of $\lambda=2200$ nm, a peak intensity of $I=10^{14}$ W/cm$^2$, and a duration of 8 cycles (total duration $\sim 60$ fs, full width at half maximum (FWHM) of intensity $\sim 20$ fs). Variation of the laser parameters will be discussed below. For this pulse shape tunneling ionization is strongly concentrated near the local field maxima of three adjacent half-cycle pulses at the center of the pulse. The final distribution in the energy ($E$) -- angular momentum ($L$) plane (Fig.\ \ref{fig1}) of electrons emitted into a double cone oriented along the polarization axis ($0^\circ$ and $180^\circ$) with opening angle $\theta_c=\pm 10^\circ$ features an island with three distinct peaks at high angular momenta.
Upon projection onto the energy axis, this high-$L$--low-$E$ island can be identified as the source of the LES with an energy of $E\approx 3$ eV. This island is well-separated from the ``background'' of low-energy low-angular momentum electrons. The sharp parabolic boundary of the latter is given by
\begin{equation}
E=\frac{L^2}{2\alpha^2\sin^2\theta_c}\, .\label{eq3}
\end{equation}
The positions ($E_i,L_i$) of the peaks within the island are determined by the variation of the subsequent field maxima within the pulse envelope. Note that they are not related to the peak sequence due to multiple reversals in a monochromatic (constant amplitude) field identified in the 1D analysis \cite{kast11}. A full quantum simulation for identical laser parameters (Fig.\ \ref{fig2}) confirms the appearance of the high-$L$ island. Note that because of the Heisenberg uncertainty relation $\langle\Delta L\Delta\theta\rangle > 1$ a straight-forward comparison between classical and quantum $E-L$ distributions is possible only for the angle-integrated emission.

\begin{figure}
\centerline{\epsfig{file=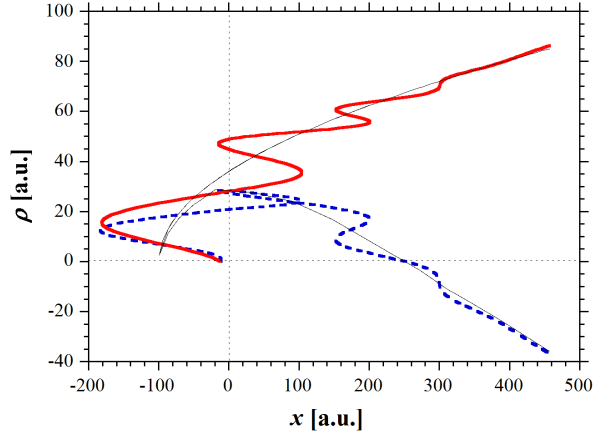,width=8cm}}
\caption{(Color online) Two typical trajectories contributing to the LES launched during the same half-cycle and eventually locked onto Kepler hyperbolae with the nucleus in its focus (thin lines). When the transverse momentum transfer near the turning point exceeds the initial transverse momentum the orientation of the Kepler hyperbola flips.}
\label{fig3}
\end{figure}

It is now instructive to explore within classical dynamics to the origin of the high-$L$--low-$E$ island. Two prototypical trajectories reaching this island (Fig.\ \ref{fig3}) visualize the strong-field dynamics perturbed by the atomic field (in the present example: a pure Coulomb field). Both trajectories are launched at the tunnel exit ($z_0\approx -10$ a.u.) near the same field maximum with slightly different transverse momenta. The subsequent quiver motion is perturbed by the combined Coulomb and laser fields upon return to the vicinity of the ionic core. The trajectories eventually lock on to Kepler hyperbolae \cite{arbo06} of different orientation but similarly large angular momenta. The rotation of the major axis of the Kepler hyperbola leading to emission into the lower hemisphere (dashed blue in Fig.\ \ref{fig3}) is consistent with the semiclassical trajectories identified in \cite{yan10}.
We emphasize that such an axis rotation is one pathway but not necessarily a prerequisite for reaching the LES as illustrated by the trajectory (solid red line in Fig.\ \ref{fig3}) emitted into the same hemisphere as originally launched. The Coulomb field provides a ``kick'' when the slowing down of the quiver motion near its turning point occurs in the vicinity of the ionic core. Such a scenario of the strong-field quiver motion only locally perturbed near the ``inner'' turning points requires a large quiver amplitude $\alpha$ compared to the typical distance from the core, taken to be of the order of the tunnel exit $z_0$,
\begin{equation}
\frac{\alpha}{z_0}\approx\frac{F_0^2}{I_p(2\pi c)^2}\lambda^2\approx\frac{2}{\gamma^2}\gg 1  \label{eq4}
\end{equation}
and readily explains why a pronounced LES develops only for mid-infrared pulses. The control parameter (Eq.\ \ref{eq4}) $\alpha/z_0=2/\gamma^2$ yields the experimentally observed scaling of the upper border of the LES, $E_H\sim\gamma^{-1.8}$, and accounts for the overall variation of the LES with $\gamma$ (but not with $I$ and $\omega$ separately) \cite{blaga09}.

\begin{figure}
\centerline{\epsfig{file=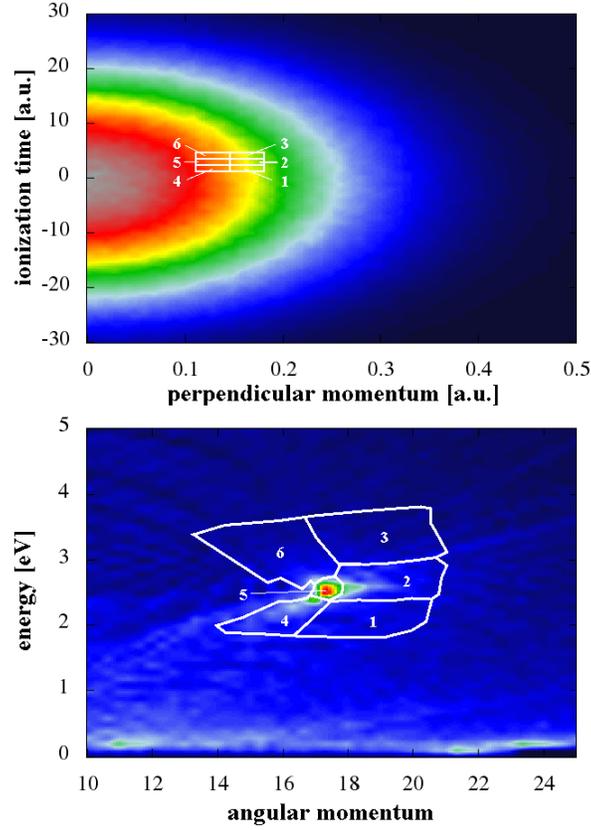,width=8cm}}
\caption{(Color online) Visualization of the Jacobian of the mapping $(\varphi,p_\perp)\to (E,L)$ for initial conditions near the absolute field maximum of a cosine-pulse at $t=0$ a.u. Shown is the dynamical mapping of the plaquettes ($\Delta \varphi_j,\Delta p_{\perp,j}$) $j=1\dots 6$ onto the $E-L$ plane. Contours indicate the initial ADK ionization probability (top) and $P(E,L)$ (bottom). To illustrate 2-dimensional mapping the longitudinal momentum is set to $p_\|=0$ a.u.\ and the position of the tunnel exit to $z_0=7$ a.u.}
\label{fig4}
\end{figure}

The formation of the high-$L$ island for the CTMC ensemble is the signature of a two-dimensional focusing in phase space. Key quantity is the classical transition probability \cite{mill74} for mapping the initial conditions, the time $t$ (or laser phase $\varphi=\omega t$), and the initial transverse momentum $p_\perp$ with which the trajectory is launched onto the final-state variables $E$ and $L$ given by the Jacobian of the dynamical transformation
\begin{equation}
P(E,L)=\left|\frac{\partial (\varphi,p_\perp)}{\partial(E,L)}\right|\, .  \label{eq5}
\end{equation}
In Eq.\ \ref{eq5} the range of $p_\perp$ is given by the ADK transverse momentum distribution and $\varphi$ is the phase angle of the laser field relative to a given field extremum $F_{0,i}$ close to which the launch of the trajectory occurs. The phase space focusing can be directly visualized by the distortion (stretching and compression) of the plaquettes $(\Delta\varphi_j,\Delta p_{\perp,j})$ of initial combinations when mapped onto the $E-L$ plane (Fig.\ \ref{fig4}).
In close analogy to pulse-induced focusing of Rydberg wavepackets \cite{arbo03}, focusing in the $E-L$ plane proceeds via the confluence of two extrema in the deflection function to a point (line) of inflection. Traces of this 2D focusing are present also in the projection onto the energy axis. This explains the appearance of the LES feature and of focusing also in reduced-dimensional models \cite{kast11}. The position of the resulting peak $(E_i,L_i)$ can be estimated from
\begin{eqnarray}
E_i&=&\frac{1}{2}\left(\frac{2F_{0,i+1}}{3\pi\omega}\right)^2\propto\frac{I_p}{\gamma^2}\label{eq6}\\
L_i&=&p_{\perp,i}\cdot\frac{F_{0,i+1}}{\omega^2}\approx p_{\perp,i}\alpha_{i+1}\, ,\label{eq7}
\end{eqnarray}
where Eq.\ \ref{eq6} follows from the impulsive momentum transfer during the longitudinal velocity reversal at the turning point close to the nucleus in the (first) subsequent half-cycle while Eq.\ \ref{eq7} follows from the locking of the quiver motion onto the asymptotic Kepler hyperbola \cite{arbo06}. Note that the final coordinates $(E_i,L_i)$ for electrons emitted during the ionization burst near the $i$-th field maximum are primarily determined by the field strength of the subsequent field maximum $F_{0,i+1}$. This observation explains the temporal order of the peaks within the island in $E$ and $L$ due to three subsequent ionization events (see Fig.\ \ref{fig1}). A further consequence is that for longer pulses with a slowly varying envelope the structure of individual peaks within the island merge into a ridge whose upper border in both $E$ and $L$ is determined by the absolute field maximum at the center of the pulse.

\begin{figure}
\centerline{\epsfig{file=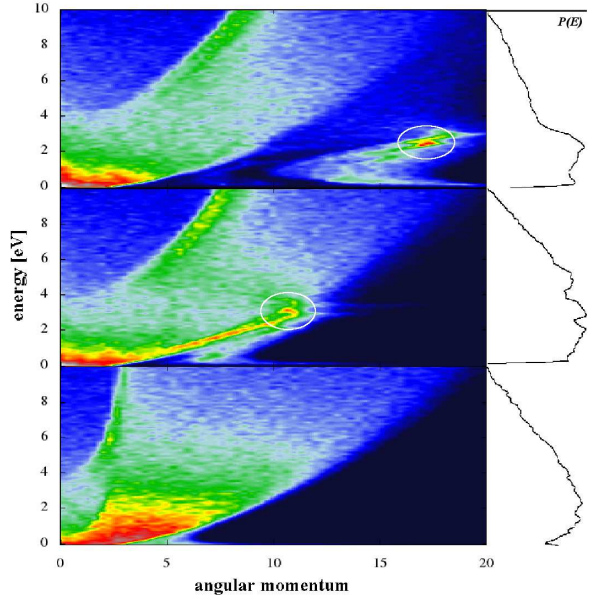,width=8cm}}
\caption{(Color online) Variation of the angular integral $E-L$ distribution with screening length $d$: top: $d=\infty , Z=1$ (Coulomb), center: $d=10, Z=1.1$, bottom: $d=0.9, Z=2$. Laser parameters as in Fig.\ \ref{fig1}, observation angle $\theta_c=\pm 10^\circ$. The resulting electron spectra are shown as projection onto the energy axis.}
\label{fig5}
\end{figure}

As the formation of the high-$L$ island results from the interplay between the strong-field dynamics and the perturbation by the atomic force field near the inner turning point, the dependence on the range of the atomic potential is of conceptual interest. We therefore employ an exponentially screened potential
\begin{equation}
V_d(r)=-\frac{Z(d)}{r}e^{-r/d},\label{eq8}
\end{equation}
where $d$ is the screening length of the potential and $Z(d)$ is adjusted such that the ionization potential remains constant at its hydrogenic value ($I_p=0.5$ a.u.). The Coulomb limit corresponds to $d=\infty$ and $Z(\infty )=1$. With decreasing $d$ the high-$L$ structure moves for $d=10$ towards lower $L\sim 10$ and disappears entirely for a truly short-ranged potential with $d\approx 1$ (Fig.\ \ref{fig5}).
Obviously, the formation of the LES-generating island requires atomic potentials whose range $d$ extends to at least to a distance of the order of the tunnel exit ($z_0\lesssim d < \alpha; z_0\approx 10$ a.u. in the present case) such that the atomic field can impart a longitudinal momentum ``kick'' to accelerate and a transversal ``kick'' to deflect the outgoing trajectory at the inner turning point of the subsequent half-cycle. In agreement with \cite{blaga09} we find a pure Coulomb field ($d=\infty$) is thus not necessarily a prerequisite for the occurrence of the LES, the shape and position of the high-$L$--low-$E$ island forming the LES, however, is very sensitive to the interactions at intermediate to large distances.

\begin{figure}[ht]
\centerline{\epsfig{file=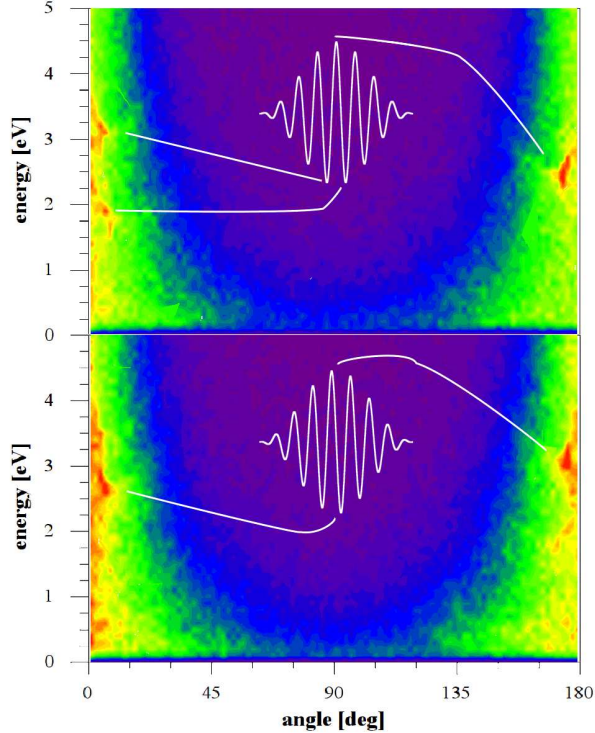,width=8cm}}
\caption{(Color online) Angle-resolved energy spectra of photoelectrons emitted in the interaction of hydrogen with an 8 cycle laser pulse with $\lambda=2200$ nm and $I=10^{14}$ W/cm$^2$. Top: cosine-pulse ($\phi_{CEP}=0$), bottom: sine-pulse ($\phi_{CEP}=\pi/2$)}
\label{fig6}
\end{figure}

The focusing effect, identified here as underlying the LES, opens up novel opportunities to monitor the carrier-envelope phase of ultrashort few-cycle mid-infrared pulses. As trajectories asymptotically propagate in the direction opposite of that of the tunnel exit, we expect different spectra in forward (defined by the direction of the maximum laser field) and backward directions (Fig.\ \ref{fig6}).
For a short cosine-pulse (here 8 cycles; $\phi_{CEP}=0$) the LES in both position and shape differs significantly from the sine-shaped pulse ($\phi_{CEP}=\pi/2$, bottom panel). For the cosine-pulse, the emission at the absolute field maximum leads to a single peak near $E=2.5$ eV and $\theta\approx 180^\circ$ observation angle while the two adjacent local maxima give a double peak structure at $E=2$ and $E=3$ eV at $\theta\approx 0^\circ$. Such differences between forward and backward directions can also be observed for other few-cycle pulses. If the carrier-envelope phase $\phi_{CEP}$ of the laser field is shifted by $\pi/2$ (8 cycles, sine-pulse, bottom panel) both forward and backward LES consist of only one peak. Due to the energy shift of the LES per half-cycle, these LES are observed at different energies (forward $\sim 2.6$ eV, backward $\sim 3$ eV). This promises to give access to information on the CEP-phase $\phi_{CEP}$. To our knowledge, this is the first identification of signatures of carrier-envelope phase effects in the low-energy part of the photoelectron spectrum. Previous studies were focused on the high-energy tail of the spectrum and shorter wavelengths ($\lambda\approx 800$ nm, \cite{paulus03}).

In conclusion, we have shown that the recently discovered low-energy structure in strong-field ionization results from a two-dimensional focusing effect of the classical phase-space distribution at high angular momenta and low energy. Full 3D quantum simulations confirm this scenario. This structure appears for small Keldysh parameters $\gamma$ or, equivalently, large ratios of quiver amplitude $\alpha$ to distances of the tunnel exit $z_0$, $\alpha/z_0\gg 1$. In this regime, the strong-field quiver dynamics is locally perturbed by the atomic force field near the ``inner'' turning point during the half-cycles subsequent to the one causing tunneling ionization. While the position and shape of the peak in the $E-L$ plane is sensitively influenced by the long-range Coulomb interaction, the appearance of an enhancement occurs also for a screened atomic force field provided its range is close to or exceeds $z_0$. We show the the LES for few-cycle pulses is sensitive to the carrier-envelope phase which may open the opportunity to measure the CEP in low-energy spectra.

This work was supported by the Austrian Science Foundation FWF under Proj.\ No.\ SFB016-ADLIS and P23359-N16 and the Elise-Richter program (S. Gr\"afe) under Proj.\ No.\ V193-N16. One of the authors (K. Dimitriou) gratefully acknowledges fruitful discussions with Th.\ Mercouris.

\end{document}